\documentclass[10pt]{llncs}
\usepackage{epic,eepic,amsmath,latexsym,amssymb,color}
\usepackage{ifthen,graphics,epsfig,subfigure}
\usepackage[english]{babel}
\usepackage{times}

\usepackage{array}
\usepackage{color}
\usepackage{colortbl}

\usepackage{pgf}
\usepackage{tikz}
\usetikzlibrary{arrows,matrix, calc, fit}

\usepackage{rotating}

\usepackage{pgfplots}

\usepackage{comment}

\usepackage{amsmath}
\usepackage{amsmath}
\usetikzlibrary{patterns}

\usepackage{floatrow}
\begin{document}

  \def\startCirc#1{\tikz[remember picture,overlay]\path node[inner sep=0, anchor=south] (st) {#1} coordinate (start) at (st.center);}%
  \def\endCirc#1{\tikz[remember picture,overlay]\path node[inner sep=0, anchor=south] (en) {#1} coordinate (end) at (en.center);%
    \begin{tikzpicture}[overlay, remember picture]%
      \path (start);%
      \pgfgetlastxy{\startx}{\starty}%
      \path (end);%
      \pgfgetlastxy{\endx}{\endy}%
      \pgfmathsetlengthmacro{\xdiff}{\endx-\startx}%
      \pgfmathsetlengthmacro{\ydiff}{\endy-\starty}%
      \pgfmathtruncatemacro{\xdifft}{\xdiff}%
      \pgfmathsetmacro{\xdiffFixed}{ifthenelse(equal(\xdifft,0),1,\xdiff)}%
      \pgfmathsetmacro{\angle}{ifthenelse(equal(\xdiffFixed,1),90,atan(\ydiff/\xdiffFixed))}%
      \pgfmathsetlengthmacro{\xydiff}{sqrt(abs(\xdiff^2) + abs(\ydiff^2))}%
      \path node[draw,rectangle,red, rounded corners=2mm, rotate=\angle, minimum width=\xydiff+4ex, minimum height=2.5ex] at ($(start)!.5!(end)$) {};%
    \end{tikzpicture}%
  }

\newlength {\squarewidth}
\renewenvironment {square}
{
\setlength {\squarewidth} {\linewidth}
\addtolength {\squarewidth} {-12pt}
\renewcommand{\baselinestretch}{0.75} \footnotesize
\begin {center}
\begin {tabular} {|c|} \hline
\begin {minipage} {\squarewidth}
\medskip
}{
\end {minipage}
\\ \hline
\end{tabular}
\end{center}
}



\newcommand{\toto}{xxx}
\newenvironment{proofT}{\noindent{\bf
Proof }} {\hspace*{\fill}$\Box_{Theorem~\ref{\toto}}$\par\vspace{3mm}}
\newenvironment{proofL}{\noindent{\bf
Proof }} {\hspace*{\fill}$\Box_{Lemma~\ref{\toto}}$\par\vspace{3mm}}
\newenvironment{proofC}{\noindent{\bf
Proof }} {\hspace*{\fill}$\Box_{Corollary~\ref{\toto}}$\par\vspace{3mm}}

\newcounter{linecounter}
\newcommand{\linenumbering}{\ifthenelse{\value{linecounter}<10}
{(0\arabic{linecounter})}{(\arabic{linecounter})}}
\renewcommand{\line}[1]{\refstepcounter{linecounter}\label{#1}\linenumbering}
\newcommand{\resetline}[1]{\setcounter{linecounter}{0}#1}
\renewcommand{\thelinecounter}{\ifnum \value{linecounter} >
9\else 0\fi \arabic{linecounter}}

\renewcommand{\note}[1]{\noindent\textcolor{red}{{\fontfamily{phv}\selectfont NOTE: #1}}}
\newcommand{\SB}[1]{\noindent\textcolor{red}{{\fontfamily{phv}\selectfont SB-NOTE: #1}}}
\newcommand{\AD}[1]{\noindent\textcolor{blue}{{\fontfamily{phv}\selectfont AD-NOTE: #1}}}
\newcommand{\MP}[1]{\noindent\textcolor{green}{{\fontfamily{phv}\selectfont MP-NOTE: #1}}}
\newcommand{\vir}[1]{``#1''}


\title{\bf Tight Mobile Byzantine Tolerant Atomic Storage}
\author{Silvia Bonomi$^\star$, Antonella Del Pozzo$^\star$, Maria Potop-Butucaru$^\dagger$\\~\\
$^\star$Sapienza Universit\`{a} di Roma,Via Ariosto 25, 00185 Roma, Italy\\
\texttt{\{bonomi, delpozzo\}}$@$dis.uniroma1.it\\
$^\dagger$Universit\'e Pierre \& Marie Curie (UPMC) -- Paris 6, France\\
maria.potop-butucaru$@$lip6.fr}
\institute{}
\date{}
\maketitle
\thispagestyle{empty}


\begin{abstract}
This paper proposes the first implementation of an atomic storage tolerant to mobile Byzantine agents. Our implementation is designed for the round-based synchronous model where the set of  
Byzantine nodes changes from round to round. In this model we explore the feasibility of \emph{multi-writer multi-reader} atomic register prone to various mobile Byzantine behaviors. We prove upper and lower bounds for solving the atomic storage in all the explored models. 
Our results, significantly different from the static case,  advocate for a deeper study of the main building blocks of distributed computing  while the system is prone to mobile Byzantine failures.
 

{\bf This paper is eligible for the Best Student Paper Award as Antonella Del Pozzo is a full time student.}\\

~\\
{\bf Type:} REGULAR PAPER\\

\end{abstract}



\section{Introduction}\label{sec:related}
Byzantine-tolerant 
storage is an active research area and this problem has been studied in various settings and models
 (e.g.\cite{B00,S90,MR98,MAD02-2} to cite just few of them).  Recently, several works investigate this problem in the case where the system starts in an arbitrary state. To cope with this situation stabilizing Byzantine tolerant algorithms have been proposed in~\cite{AADDPT15,BPT15,DDPT12}. 
 In all the above mentioned works the set of Byzantine processes is assumed to be static. 
That is, the set of nodes exhibiting a Byzantine behavior does not change during the computation. 

In the current work we investigate a different fault model where Byzantines are mobile. This model captures insiders attacks or viruses propagation. In the mobile Byzantine fault model transient state corruptions, which can be abstracted as Byzantine ``agents,'' can move through the network and corrupt the nodes they occupy. A node occupied by a Byzantine agent will behave arbitrarily for a transient period of time.  Once the Byzantine agent leaves the node, the node eventually behaves correctly. However, the Byzantine agent may "infect" another node that behaved correctly until the infection. This models the situation where, as soon as a faulty node is repaired, another one becomes compromised.
 
There are two main research directions in the mobile Byzantine area: Byzantines with constrained mobility and Byzantines with unconstrained mobility. In both models the only distributed problem studied so far is the  agreement problem. Byzantines with constraint mobility were studied by Buhrman \emph{et al.} \cite{Garay+95+ORA}. They consider that Byzantine agents move from one node to another only when protocol messages are sent (similar to how viruses would propagate). 

In the case of unconstrained mobility the motion of Byzantine agents is not tight to the message exchange. Several authors investigated the agreement problem in variants of this model: \cite{Banu+2012,BDNP14,Garay+1994,Ostrovsky+91,Reischuk+85,Sasaki+2013}.
Reischuk \cite{Reischuk+85}  investigate the stability/stationarity of malicious agents for a given period of time.  Ostrovsky and Yung \cite{Ostrovsky+91} introduced the notion of mobile virus and investigate an adversary that can inject and distribute faults. 

Our work follows the lines opened by Garay \cite{Garay+1994}.
Garay \cite{Garay+1994} and, more recently, Banu \emph{et al.} \cite{Banu+2012} and Sasaki  \emph{et al.}  \cite{Sasaki+2013} or Bonnet  \emph{et al.} \cite{BDNP14} consider, in theirs models, that processes execute synchronous rounds composed of three phases: \emph{send}, \emph{receive}, \emph{compute}. Between two consecutive rounds, Byzantine agents can move from one host to another, hence the set of faulty processes has a bounded size although its membership can change from one round to the next.  
 
 In the current work we focus four of the above discussed models, all four consider a synchronous round-based system : Garay \cite{Garay+1994},  Buhrman \emph{et al.} \cite{Garay+95+ORA}, Sasaki  \emph{et al.}  \cite{Sasaki+2013} and Bonnet  \emph{et al.} \cite{BDNP14}. 
In the Garay's model  a process has the ability to detect its own infection after the Byzantine agent left it.
More precisely, during the first round following the leave of the Byzantine agent, a process enters a state, called \emph{cured}, during which it can take preventive actions to avoid sending messages that are based on a corrupted state.   Garay \cite{Garay+1994} proposes in this model an algorithm that solves Mobile Byzantine Agreement provided that $n>6t$ (dropped later to $n>4f$ in \cite{Banu+2012}). 

Buhrman \emph{et al.} \cite{Garay+95+ORA} propose a model where the motion of Byzantine agents is tight to the message exchange. In this model they prove a tight bound for Mobile Byzantine Agreement ($n > 3t$, where $t$ is the maximal number of simultaneously faulty processes) and propose a time optimal protocol that matches this bound. 

Bonnet  \emph{et al.} \cite{BDNP14}  investigated the same problem in a  model where processes do not have the ability to detect when Byzantine agents move. However, differently from Sasaki  \emph{et al.}  \cite{Sasaki+2013}, cured processes have \emph{control} on the messages they send. This subtle difference on the power of Byzantine agents has an impact on the bounds for solving the agreement. If in the Sasaki's model the bound on solving agreement is $n>6f$  in Bonnet's model it is $n>5f$ and this bound is proven tight.

 \paragraph{Our contribution.}
 As far as we known, our construction is the first that builds a distributed MWMR
atomic memory on top of synchronous round-based servers, which communicate by 
message-passing,  and where some of them can 
 exhibit a Byzantine behavior induced by a mobile malicious agent. We prove first upper bounds on the number of faulty processes for four of the mobile Byzantine models cited above: Garay \cite{Garay+1994}, Buhrman \emph{et al.} \cite{Garay+95+ORA}, Sasaki  \emph{et al.}  \cite{Sasaki+2013} and Bonnet  \emph{et al.} \cite{BDNP14}.
Then, we propose tight implementations of a atomic register in each of these models altogether with their correctness proofs.
The first study focuses the model of Garay {\it et al.} \cite{Garay+1994}, where nodes can detect that they were previously infected by a Byzantine agent and remain silent until their state is cleaned. In this model, we implement the atomic register provided that in each round the number of Byzantine nodes (nodes occupied by a Byzantine agent), $f$, is less than $n/3$ where $n$ is the number of correct nodes in that round. The second study concerns the models of Sasaki  \emph{et al.}  \cite{Sasaki+2013} and Bonnet  {\it et al.} \cite{BDNP14}, where infected nodes cannot locally detect the presence or the absence of a Byzantine agent and hence can send/compute based on a corrupted state even thought the mobile agent is not anymore located at that node. In both these models we implement the atomic register provided that in each round the number of Byzantine nodes $f$ is less than $n/4$ where $n$ is the number of correct nodes in the round.  Note that differently from the case of the agreement problem, these models have the same power in the case of atomic memory implementation. The last studied model is Buhrman \emph{et al.} \cite{Garay+95+ORA} where Byzantine agents move with the messages. In this model, we provide an implementation of the atomic memory provided that $f$ is less than $n/2$.
Note that all the above bounds are also lower bounds for the considered models.  
 
 \paragraph{Paper roadmap.}
 The paper is organized as follows. In Section \ref{sec:Model} we define the model of the system and the problem of MWMR atomic memory.
In  Section \ref{sec:upper-bounds} we prove upper bounds on the faulty processes necessary to implement MWMR atomic memory in the following four mobile Byzantine models: Garay \cite{Garay+1994}, Buhrman \emph{et al.} \cite{Garay+95+ORA}, Sasaki  \emph{et al.}  \cite{Sasaki+2013} and Bonnet  \emph{et al.} \cite{BDNP14}.
 In Section \ref{sec:algorithm}  we present a generic tight algorithm that implements MWMR atomic memory parametrized function on the considered mobile Byzantine model. The correctness of the generic algorithm is proved in Section \ref{sec:proofs}. Finally, Section \ref{sec:conclusions} concludes the paper and discuss some open research directions.

\section{Model and Problem Definition}
\label{sec:Model}
\subsection{System Model}
\label{sec:systemModel}


We consider a distributed system composed of an arbitrary large set of clients $\mathcal{C}$ and a set of $n$ servers $\mathcal{S}=\{s_1, s_2 \dots s_n\}$. Each process in the distributed system (i.e., both servers and clients) is identified trough a unique integer identifier.
Servers run a distributed protocol implementing a shared memory abstraction.\\  
 
\noindent {\bf Communication model and timing assumptions.} Processes communicate trough message passing. 
In particular, we assume that (i) each client $c_i \in \mathcal{C}$ can communicate with every server trough a ${\sf broadcast}$ primitive, (ii) servers can communicate among them trough a ${\sf broadcast}$ primitive and (iii) servers can communicate with clients trough point-to-point channels.
We assume that communications  are authenticated (i.e., given a message $m$, the identity of its sender cannot be forged) and reliable (i.e. messages are not created, lost or duplicated).

The system evolves in synchronous rounds. Every round is divided in three phases: (i) \emph{send} where processes send all the messages for the current round, (ii) \emph{receive} where processes receive all the messages sent at the beginning of the current round and (iii) \emph{computation} where processes process received messages and prepare those that will be sent in the next round. 
Processes have access to the current round number via a local variable that we usually denote by $r$.\\ 

\noindent{\bf Failure model.} We assume that an arbitrary number of clients may crash while servers are affected by \emph{mobile Byzantine failures} (MBF) \cite{BDNP14,Garay+1994,Garay+95+ORA,Sasaki+2013}.
Informally, in the mobile Byzantine failure model, faults are represented by powerful computationally unbounded agents that move arbitrarily from a server to another. When the agent is on the server, it can corrupt its local variables, force it to send arbitrary messages (potentially different from process to process) etc... However, the agent cannot corrupt the identity of the server.
We assume that, in each round, at most $f$ servers can be affected by a mobile Byzantine failure. 
When an agent occupies a server $s_i$ we will say that $s_i$ is {\em faulty}. 
When the agent leaves $s_i$ it is said to be {\em cured} until it does not restore the correct internal state. If a server is neither {\em faulty} nor {\em cured} then it is said to be {\em correct}. 
We assume similar to \cite{BDNP14,Garay+1994,Sasaki+2013} that each server has a tamper-proof memory where it safely stores the correct algorithm code.
When the agent leaves a server $s_i$ (i.e., it becomes {\em cured}), it recovers the correct algorithm code from the tamper-proof memory.
Concerning the assumptions on agent movements and the server awareness on its {\em cured} state, different models have been defined. In the paper we will consider all the variants of mobile Byzantine failures \cite{BDNP14,Garay+1994,Garay+95+ORA,Sasaki+2013}:

\begin{itemize}
	\item {\bf (M1)} \emph{Garay's model} \cite{Garay+1994}. In this model, agents can move arbitrarily from a server to another at the beginning of each round (i.e. before the send phase starts). When a server is in the $cured$ state it is aware of its condition and thus can remain silent to prevent the dissemination of wrong information until its code has been completely restored and its state is corrected. 

	\item {\bf (M2)} \emph{Bonnet et al.'s model} \cite{BDNP14} and {\bf (M3)} \emph{Sasaki et al.'s model} \cite{Sasaki+2013}. As in the previous model, agents can move arbitrarily from a server to another at the beginning of each round (i.e. before the send phase starts). Differently from the Garay's model, in both models it is assumed that servers do not know if they are correct or cured when the Byzantine agent moved. The main difference between these two models is that in the \cite{Sasaki+2013} model a cured process still acts as a Byzantine one extra round.  
	
	\item {\bf (M4)} \emph{Buhrman's model} \cite{Garay+95+ORA}. Differently from the previous models, agents move together with the message (i.e., with the ${\sf send}$ or ${\sf broadcast}$ operation). However, when a server is in the {\em cured} state it is aware of that. 
	
\end{itemize}

\subsection{Atomic Registers}
\label{sec:Problem}

A register is a shared variable  accessed by a set of processes, i.e. clients,  through two operations, namely ${\sf read()}$ and ${\sf write()}$. Informally, the ${\sf  write()}$ operation updates the value  stored in the shared variable while  the $\sf read()$ obtains the  value contained in the variable (i.e. the last written value). Every operation issued on a register is, generally, not instantaneous and it  can  be characterized  by  two events  occurring  at  its boundary:  an \emph{invocation} event and a \emph{reply} event. These events occur at two time instants (invocation  time and reply time) according  to the fictional global time.\\  
An operation $op$  is \emph{complete} if both the  invocation event and the reply event occur (i.e. the  process executing the operation does not crash between the invocation and the reply).
Contrary, an operation $op$ is said to be \emph{failed} if it is invoked by a process that crashes before the reply event occurs. According to these time instants, it is possible to state when two operations are concurrent with respect to the real time execution.
For ease of presentation we assume the existence of a fictional global clock and the invocation time and response time of every operation are defined with respect to this fictional clock.\\
Given two operations  $op$ and $op'$, and their  invocation event and reply event  times  ($t_{B}(op)$ and $t_B(op')$) and return  times ($t_E(op)$ and $t_E(op')$),  we say that $op$ \emph{precedes} $op'$ ($op \prec op'$) iff $t_E(op) < t_B(op')$. If $op$ does not precede $op'$ and $op'$  does not  precede $op$,  then $op$  and $op'$   are \emph{concurrent} ($op||op'$). 
Given a  ${\sf write}(v)$ operation,  the value $v$  is said to  be written when the operation  is complete.\\ 
We assume that locally any client never performs ${\sf read()}$ and ${\sf write()}$ operation concurrently. We also assume that initially the register stores a default value $\bot$ written by a fictional ${\sf write}(\bot)$ operation happening instantaneously at round $r_0$. 
In case of concurrency while  accessing the shared variable, the meaning of \emph{last written  value} becomes ambiguous.  Depending  on the  semantics of the operations, three types of register   have   been   defined   by  Lamport \cite{L86}:   \emph{safe}, \emph{regular} and \emph{atomic}. 
In this paper, we will consider a Multi-Writer/Multi-Reader (MWMR) atomic register which is specified as follows:
\begin{itemize}
	\item ${\sf Termination}$:
	If a correct client invokes an operation, it eventually  returns from that operation. 
	\item ${\sf Validity}$:
	A read operation returns the last value written before its invocation, or a value written by a write operation concurrent with it.
	\item ${\sf Ordering}$:
	There exists a total order $S$ of ${\sf read}()$ and ${\sf write}()$ operations such (i) if $op \prec op'$ then $op$ appears before $op'$ in $S$ and (ii) any ${\sf read}()$ operation returns the value $v$ written by the last ${\sf write}()$ preceding it in $S$.
\end{itemize}

\section{Upper Bounds on the number of Faults}

%
The next theorems provide upper bounds on the number of faulty processes for the implementation of MWMR Atomic Register in the models of mobile Byzantine faults \cite{BDNP14,Garay+1994,Garay+95+ORA,Sasaki+2013}.

\begin{theorem}\label{th:GarayTight}
	If $n\le 3f$, there exists no algorithm  that implements a MWMR Atomic Register in the Garay's model \cite{Garay+1994}. 
\end{theorem}

\begin{proofT}
Consider that each ${\sf read}()$ operation takes at least one round to be executed and, according to the Garay's model, at the beginning of each round servers are partitioned in three sets: (i) faulty, (ii) cured and (iii) correct.
Due to the assumption that we have $f$ faulty servers in each round, we have that, cured processes, in the worse case, are $f$ as well (i.e., the $f$ servers that were faulty in the previous round). Thus, considering that $n$ is at most $3f$, we follows that, in the worst case, at most $f$ processes are correct.
As a consequence, considering that cured servers are silent (they do not send any message), the reader will gather at most $2f$ values and it will be not able to distinguish those that come from correct servers from those coming from faulty one.
	\renewcommand{\toto}{th:GarayTight}
\end{proofT}

\begin{theorem}\label{th:SasakiTight}
	If $n\le 4f$, there exists no algorithm that implements a MWMR Atomic Register in the Sasaki's model \cite{Sasaki+2013}. 
\end{theorem}

\begin{proofT}
	The claim simply follows by considering that each ${\sf read}()$ operation takes at least one round to be executed and, according to the Sasaki's model, at the beginning of each round servers are partitioned in three sets: (i) faulty, (ii) cured and (iii) correct.
	Due to the assumption that at most $f$ faulty servers are in each round, it follows that, cured processes, in the worst case, are $f$ (i.e., the $f$ servers that was faulty in the previous round). Thus, considering that $n$ is at most $4f$, we have that, in the worst case, at most $2f$ processes are correct.
	As a consequence, considering that cured servers act like faulty ones as well, the reader will get back at most $4f$ values and it will be not able to distinguish which ones come from correct servers (i.e., $2f$ same values $v$) from those coming from faulty one (i.e., $2f$ same values $v'$).
	\renewcommand{\toto}{th:SasakiTight}
\end{proofT}

\begin{theorem}\label{th:BonnetTight}
	If $n\le 4f$, there exists no algorithm that implements a MWMR Atomic Register in the Bonnet's model \cite{BDNP14}.
\end{theorem}

\begin{proofT}
	The claim simply follows by considering that the Bonnet's model is a particular case of Sasaki model, in which cured servers act as less powerful faulty servers, forced to send the same message to all. The same reasoning as in the proof of Theorem \ref{th:SasakiTight}  is applied.
	\renewcommand{\toto}{th:BonnetTight}
\end{proofT}

\begin{theorem}\label{th:BurhmanTight}
	If $n\leq 2f$ there exists no algorithm that implements a MWMR Atomic Register in the Burhman's model \cite{Garay+95+ORA}. 
\end{theorem}

\begin{proofT}
The proof is similar to the static case \cite{B00}.
	Let us suppose by contradiction that such algorithm exists and suppose without restraining the generality that $n=2f$. 
	Let $v$ be the value written by the last completed ${\sf write}()$ operation and let us assume that no other operations are concurrent with the ${\sf read}()$.
In this settings, when the client gets values from servers, it will receive at most $f$ same value $v$ from correct servers and $f$ same values $v'$, with $v'\neq v$ from faulty servers.
As a consequence, the reader has no way to distinguish between the two values and we have a contradiction.

	\renewcommand{\toto}{th:BurhmanTight}
\end{proofT}

\label{sec:upper-bounds}
\section{Tight MWMR Atomic Register Implementation}\label{sec:algorithm}

In this section we present a generic algorithm $\mathcal{A}_{Areg}$ (Fig.\ref{fig:client}-\ref{fig:server}) that implements the MWMR Atomic Register in all the above presented models. In order to abstract the knowledge a server has on its state (i.e. {\em cured} or {\em correct}), we introduce the  ${\sf cured\_state}$ oracle.
When invoked via ${\sf report\_cured\_state}()$ function it returns ${\sf true}$ to {\em cured} servers and ${\sf false}$ to others in the Garay \cite{Garay+1994} and Buhrman \emph{et al.} \cite{Garay+95+ORA}. In this case the oracle is said enabled. ${\sf cured\_state}$ oracle returns always ${\sf false}$ in Sasaki  \emph{et al.}  \cite{Sasaki+2013} or Bonnet  \emph{et al.} \cite{BDNP14} models. In this case the oracle is said disabled.

  In the following we propose a generic MWMR atomic register algorithm that is tight for all the above models by just tuning the following three parameters: $\alpha$, $\beta$ and the ${\sf cured\_state}$ oracle status. Let denote the number of servers with respect to {\em faulty} servers by $n>\alpha f$, where $\alpha \in \{2,3,4\}$ following the mobile Byzantine model. Let $s$ be the minimal number of required occurrences of the same value in order to chose it, $s= n-\beta f$. Basically $s$ has to be greater than the number of possible wrong values that {\em faulty} and {\em cured} servers can return, which is $\beta f$, where $\beta \in \{1,2\}$ depending on the model adopted for the {\em cured} servers.

Table \ref{t:parameters} summarizes the above in a synthetic way. 

\begin{table}
\begin{tabular}{| c | c | c | c | c |}
\hline
Failure model& M${id}$ & $\alpha$ & $\beta$ & Oracle\\
\hline
Garay \cite{Garay+1994}& M1 & 3& 2 & enabled \\
Bonnet \emph{et al.} \cite{BDNP14}& M2 & 4& 2 & disabled \\
Sasaki \emph{et al.}  \cite{Sasaki+2013}& M3 & 4& 2 & disabled \\
Burhman \emph{et al.} \cite{Garay+95+ORA}& M4 & 2& 1 & enabled\\
\hline
\end{tabular}
\caption{$\mathcal{A}_{Areg}$ parameters for the four different Mobile Byzantine Failure models.}
\label{t:parameters}
\end{table}

\begin{figure}[t]
	\centering
	\fbox{
		\begin{minipage}{0.4\textwidth}
			\scriptsize
			\resetline
			\begin{tabbing}
				aaaA\=aA\=aA\=aA\=aaaA\kill\\
				{\bf At the beginning of each round $r$}\\
				
				\line{s1}\> $echo\_vals_i \leftarrow \emptyset$;\\
				\line{s2}\> $current\_writes_i \leftarrow \emptyset$;\\
				\line{s3}\> $cured_i \leftarrow {\sf report\_cured\_state}()$;\\ 	
				
				-----------------------------------------------\\
				
				{\bf Send Phase of round $r$}\\
				
				\line{s4}\>{\bf if} \= $(\neg cured_i)$\\
				\line{s5}\>\> {\bf then} \= ${\sf broadcast}$ {\sc echo}$(val, i)$; \hspace{2.3cm} $\%$ maintenance\\ 
				\line{s6}\>\>\> {\bf for} \= {\bf each} $j \in current\_reads_i$ {\bf do}\\
				\line{s7}\>\>\>\> ${\sf send}$ {\sc reply}$(value_i)$ to $c_j$; \hspace{2.0cm} $\%$ reply to ${\sf read}()$ operations started in round $r-1$\\ 			
				\line{s7}\>\>\> {\bf endFor}\\
				\line{s9}\>{\bf endif}\\
				\line{s10}\>$current\_reads_i \leftarrow \emptyset$;\\
				
				-----------------------------------------------\\
				
				{\bf Receive Phase of round $r$}\\
				
				\line {s11} \> {\bf for} \= {\bf each}  {\sc echo}$(v, j)$ message in $rcv_i$ {\bf do}\\
				\line {s12} \>\> $echo\_vals_i \leftarrow echo\_vals_i \cup{v}$;\\
				\line {s13} \> {\bf endFor}\\
				
				\line {s14} \> {\bf for} \= {\bf each}  {\sc write}$(v, j)$ message in $rcv_i$ {\bf do}\\
				\line {s15} \>\> $current\_writes_i \leftarrow current\_writes_i \cup{<v, i>}$;\\
				\line {s16} \> {\bf endFor}\\
				
				\line {s17} \> {\bf for} \= {\bf each}  {\sc read}$(j)$ message in $rcv_i$ {\bf do}\\
				\line {s18} \>\> $current\_reads_i \leftarrow current\_reads_i \cup \{j\}$;\\
				
				-----------------------------------------------\\
				
				{\bf Computation Phase of round $r$}\\
				
				\line {s19} \> {\bf if} \= $(current\_writes_i \neq \emptyset)$\\
				\line {s20} \>\> {\bf then} \= {\bf let} $v$ {\bf such that} $\exists <v, j> \in current\_writes_i \wedge j={\sf max}_{k}(<-, k>)$;\\
				\line {s21} \>\>\> $value_i \leftarrow v$;\\
				\line {s22} \>\> {\bf else} \= {\bf if} \= $(\exists v \in echo\_vals_i ~|~ {\sf \#occurrence}(v) \ge n-\beta f)$\\
				\line {s23} \>\>\>\> {\bf then} \= $value_i \leftarrow v$;\\
				\line {s24} \>\>\> {\bf endif}\\
				\line {s25} \> {\bf endif}\\				
						
			\end{tabbing}
			\normalsize
		\end{minipage}
		
	}
	
	\caption{$\mathcal{A}_{Areg}$ implementation: code executed by any server $s_i$.}
	\label{fig:server}   
\end{figure}

\begin{figure}[t]
	\centering
	
	\fbox{
		\begin{minipage}{0.4\textwidth}
			\scriptsize
			\resetline
			\begin{tabbing}
				aaaA\=aA\=aA\=aA\=aA\=aaaA\kill\\
				
				{\bf operation} ${\sf read}()$:\\
				
				\line{c1}\> $to\_send_i \leftarrow to\_send_i \cup \{$ {\sc read}$(i) \}$;\\
				\line{c2}\> $reading_i \leftarrow {\sf true}$;\\

				-----------------------------------------------\\
				
				{\bf operation} ${\sf write}(v)$\\
				
				\line{c3}\> $to\_send_i \leftarrow to\_send_i \cup \{$ {\sc write}$(v, i) \}$;\\
				\line{c4}\> $writing_i \leftarrow {\sf true}$;\\
				
%
%
				-----------------------------------------------\\
				
				{\bf Send Phase of round $r$}\\
				
				\line{c5} \> {\bf for each} {\sc m}$()$ $\in to\_send_i$ {\bf do}  ${\sf broadcast}$ {\sc m}$()$;\\
				\line {c6} \> {\bf if} ($op\_start_i == \bot$)\\
				\line {c7} \>\> {\bf then} $op\_start_i \leftarrow r$;\\
				\line {c8} \> {\bf endIf} \\
				\line {c9} \> $to\_send_i \leftarrow \emptyset$;\\
				
				-----------------------------------------------\\
				
				{\bf Receive Phase of round $r$}\\
				
				\line {c10} \> {\bf for} \= {\bf each}  {\sc reply}$(v, j)$ message in $rcv_i$ {\bf do}\\
				\line {c11} \>\> $replies_i \leftarrow replies_i \cup{<v, j>}$;\\
				\line {c12} \> {\bf endFor}\\
				
				-----------------------------------------------\\
				
				{\bf Computation Phase of round $r$}\\
				
				\line {c13} \> {\bf if} \= $(writing_i \wedge op\_start_i = r)$\\
				\line {c14} \>\> {\bf then} \= $writing_i \leftarrow {\sf false}$;\\
				\line {c15} \>\>\> $op\_start_i \leftarrow \bot$;\\
				\line {c16} \>\>\> {\bf return} ${\sf write\_confirmation}$;\\
				\line {c17} \> {\bf endif}\\
				
				\line {c18} \> {\bf if} \= $(reading_i \wedge op\_start_i = r-1)$\\
				\line {c19} \>\> {\bf then} \= $reading_i \leftarrow {\sf false}$;\\
				\line {c20} \>\>\> $op\_start_i \leftarrow \bot$;\\
				\line {c21} \>\>\> {\bf let} $v$ {\bf such that} $\exists <v, j> \in replies_i \wedge {\sf \#occurrence}(v) \ge n-\beta f$;\\
				\line {c22} \>\>\> $replies_i \leftarrow \emptyset$;\\
				\line {c23} \>\>\> {\bf return} $v$;\\
				\line {c24} \> {\bf endif}\\

			\end{tabbing}
			\normalsize
		\end{minipage}	
		
	}
	
	\caption{$\mathcal{A}_{Areg}$ implementation: code executed by any client $c_i$.}
	\label{fig:client}   
\end{figure}

\subsection{$\mathcal{A}_{Areg}$ Algorithm description}

The presented algorithm exploits the round based nature of the system model. Any ${\sf write}()$ operation lasts one round, during which a client sends the value and all servers deliver it in the same round. Due to the synchrony assumptions no acknowledgement messages are required and the operation can terminate. If more than one ${\sf write}()$ operation falls in the same round then any server receives the same set of values. The one coming from the client with the highest identifier is stored, thus any server chose the same value. The ${\sf read}()$ operation lasts two rounds. One round to send a read request to servers and the subsequent one to gather replies. The value which occurrence is at least the threshold $n-\beta f$ is returned. \\
Along with the classical ${\sf read}()$ and ${\sf write}(v)$ operations performed by clients, for maintenance purpose in each round servers echo each other their value. Thus even though at each round at most $f$ servers may lose the value (and no  ${\sf write}()$ operation occurs), thanks to the echoed values at the end of each round {\em cured} servers are able to became {\em correct}, having the same {\em correct} servers value.

\paragraph{Client local variables.}
Each client $c_i$ manages the following variables:\\
\noindent $\mathbf{-}$ $to\_send_i$: a set in which are stored messages to be sent in the next {\em send} phase and emptied just after.\\
\noindent $\mathbf{-}$ $reading_i$ and $writing_i$: two boolean variables, only the one corresponding to the current operation is set to ${\sf true}$. \\
\noindent $\mathbf{-}$ $op\_start_i$: a variable in which is stored the current round when a new operation starts and set to $\bot$ when it ends.\\ 
\noindent $\mathbf{-}$ $rcv_i$ is a set variable (emptied at the beginning of each round), where $c_i$ stores messages received during the current round r.\\
\noindent $\mathbf{-}$ $replies_i$: a set in which are stored messages delivered after a read request.

\paragraph{Server local variables.}
Each server $s_j$ manages the following variables:\\
\noindent $\mathbf{-}$ $value_i$: the maintained value.\\
\noindent $\mathbf{-}$ $rcv_i$ is a set variable (emptied at the beginning of each round), where $s_j$ stores messages received during the current round r.\\
\noindent $\mathbf{-}$ $echo\_vals_j$: a set (emptied at the beginning of each round), in which are stored the echoed values by servers in each round.\\
\noindent $\mathbf{-}$ $current\_writes_j$: a set (emptied at the beginning of each round), in which are stored values that clients want to write during the current round.\\
\noindent $\mathbf{-}$ $currend\_reads_j$: a set in which are stored the identifiers of clients whose requested for a read. It is emptied after the reply to such clients.\\
\noindent $\mathbf{-}$ $cured_j$: boolean variable set through the ${\sf report\_cured\_state}()$ event. It is set to ${\sf true}$ by the ${\sf cured\_state}$ oracle (if enabled) when $s_j$ is in a $cured$ state. Otherwise it is always ${\sf false}$.  

\paragraph{Server maintenance.}
For maintenance purposes, at the beginning of each round, servers exchange their stored value $value_j$ 
allowing {\em cured} servers to became {\em correct} at the end of it. Thus, during the {\em send} phase of each round, servers ${\sf broadcast}$ the $ECHO(val, i)$ message (Fig.\ref{fig:server}, line \ref{s5}). If not new values have been written in the current round (the condition at line \ref{s19} is not verified), during the {\em computation} phase (Fig.\ref{fig:server}, line \ref{s22}) 
they chose the one with at least $n- \beta f$ occurrences. Note that in the case in which servers are aware of being in a {\em cured} state (Fig.\ref{fig:server}, line \ref{s4}) then they avoid to send their $value_j$.

\paragraph{Write operation.} 
When a client $c_i$  wants to write a value $v$, it  stores in $to\_send_i$ a message $WRITE(v,i)$ and sets the variable $writing_i$ to ${\sf true}$ (Fig.\ref{fig:client}, line \ref{c3}-\ref{c4}). At the subsequent {\em send} phase, $c_i$ broadcasts $WRITE(v,i)$ to all servers, stores the current round in $op\_start_i$ and empties the $to\_send_i$ set (Fig.\ref{fig:client}, line \ref{c5}-\ref{c9}). At the server side this message will be delivered within the same round during the {\em receive} phase and any $correct$ and $cured$ server $s_j$ stores it in $current\_writes_j$ set (Fig.\ref{fig:server}, line \ref{s14}-\ref{s15}).
At the end of the round, during the {\em computation} phase, if $current\_writes_j$ is not empty then the value associated to the highest client identifier is stored in $value_j$ (Fig. \ref{fig:server}, line \ref{s19}-\ref{s21}).\\
Back to the client side, during its $computation$ phase if $writing_i$ is ${\sf true}$ and $op\_start_i$ is equal to the current round $r$, this means that during the current round $c_i$ performed a ${\sf write}()$ operation. Since it lasts just one round then it sets $writing_i$ to ${\sf false}$, $op\_start_i$ to $\bot$ and returns the ${\sf write\_conformation}$ to the application layer (Fig. \ref{fig:client}, line \ref{c13}-\ref{c17}).\\

\paragraph{Read operation.}
When a client $c_i$ wants to read at round $r$ then it  stores in $to\_send_i$ a message $READ(i)$ and sets the variable $reading_i$ to ${\sf true}$ (Fig.\ref{fig:client}, line \ref{c1}-\ref{c2}). At the subsequent {\em send} phase $c_i$ broadcasts a $READ(i)$ message to all servers, stores the current round $r$ in $op\_start_i$ and empties the $to\_send_i$ set (Fig.\ref{fig:client}, line \ref{c5}-\ref{c9}). Note, the check at line \ref{c6} is necessary to avoid that $op\_start_i$ would be updated at each round. This would not be an issue for the ${\sf write}()$ operation which lasts only one round, but in the case of ${\sf read}()$ operation it would cause the loss of information about the starting round. At server side, the $READ(i)$ message will be delivered within the same round $r$ and any {\em correct} and {\em cured} server $s_j$ stores the client identifier in the $current\_reads_j$ set (Fig. \ref{fig:server}, line \ref{s17}-\ref{s18}).\\ 
At the start of the next round $r+1$, 
if server $s_j$ is not {\em cured} or not aware of that then it sends the message $REPLY(value_j)$ to all the clients in $current\_reads_j$ set, which is emptied at the end of the {\em send} phase (Fig. \ref{fig:server}, line \ref{s6}-\ref{s10}). At client side all the $REPLY(value_j)$ are delivered and stored in the set $replies_i$ during the {\em receive} phase (Fig.\ref{fig:client},line \ref{c10}-\ref{c12}). Now during the {\em computation} phase the $reading_i$ variable is {\sf true} and $op\_start_i$ is storing the previous round number. Thus $reading_i$ is set to {\sf false}, $op\_start_i$ is set to $\bot$ and the value in $replies_i$ which occurs more than $n-\beta f$ times is returned to the application layer and $replies_i$ is emptied (Fig. \ref{fig:client}, line \ref{c18}-\ref{c24}).


\subsection{Correctness Proofs}\label{sec:proofs}

\begin{lemma}\label{l:consistencyEndGBS}


Let $\alpha_{Mi}$ and $\beta_{Mi}$ be the parameters for each of the $4$ failure models Mi as reported in Table \ref{t:parameters} and used by the algorithm in Fig. \ref{fig:server}-\ref{fig:client}.
Let $n > \alpha_{Mi} f$ for each failure model Mi considered. 
At the end of each round, at least $n-f$ correct servers store the same value $v$ in their $value_i$ local variable.
\end{lemma}

\begin{proofL}
Each non-faulty server updates its $value_i$ local variable at the end of each round $r$ (i) in line  \ref{s21} i.e., if there exists at least a pair in the $current\_writes_i$ local variable, or (ii) in line  \ref{s23} i.e., $current\_writes_i$ is empty and there exist at least $n-\beta f$ same values in $echo\_vals_i$.

First we prove that one of the two cases always happens and then we prove that the number of non-faulty servers storing the same values $v$ is $n-f$.
The $current\_writes_i$ local variable is initialized by any non-faulty server $s_i$ to $\emptyset$ at the beginning of each round $r$ (cfr. line \ref{s2}) and it is updated when a {\sc write}$()$ message is received by $s_i$\footnote{Recall that such {\sc write}$()$ message is sent by the writer client in the send phase of the first round starting after the ${\sf write}()$ invocation and it is delivered by any non-faulty server in the same round.}.
Thus, case (i) corresponds to a scenario where at least a ${\sf write}()$ operation is executed in round $r$ and case (ii) corresponds to a scenario where no ${\sf write}()$ is running.

\begin{itemize}
\item {\bf Case (i):} $\mathbf{current\_writes_i \neq \emptyset.}$ In this case the claim simply follows by considering that (i) writer clients broadcast a {\sc write}$(v, j)$ message in the send phase of round $r$, (ii) clients are correct so the same set of values is delivered to all servers that will apply a deterministic function to select the value $v$ and (iii) at most $f$ servers are faulty and may skip the update of their $value_i$ variable.\\

\item {\bf Case (ii):} $\mathbf{current\_writes_i = \emptyset}$ {\bf and line \ref{s22} is true.} In this case, the $value_i$ variable is updated according to the values stored in $echo\_vals_i$. Such variable is emptied by every non-faulty process at the beginning of each round (cfr. line \ref{s1}) and is filled in when an {\sc echo}$()$ message is delivered.
Such message is sent at least by any server, believing it is correct, at the beginning of each round.
Let $r'$ be the round in which the last ${\sf write}(v)$ operation terminated. Note that, due to above hypothesis, a ${\sf write}()$ operation always exists as we assume a fictional write happening instantaneously at round $r_0$. Without loss of generality, let us consider the round $r=r'+1$.
Due to case (i), at the end of $r'$, at least $n-f$ non-faulty servers store the same value $v$ in their local variable $value_i$.
Thus, at the beginning of $r'+1$, at least $n-f-x$ correct servers will send an {\sc echo}$(v, j)$ message, where $x$ is the number of non-faulty processes that become faulty while passing from $r'$ to $r$ (i.e. $x=f$ for all the models but Burhman's one where $x=0$ as faulty processes move during the send phase and not at the beginning of the round).
It follows that the condition in line \ref{s22} is verified if and only if $n-f-x \ge n-\beta f$ that is true in any model.
Therefore, considering that at the end of round $r$ non-faulty servers are exactly $n-f$, we have that $n-f$ processes will execute this update.
Iterating the reasoning for any $r$ the claim follows.

\end{itemize}
	\renewcommand{\toto}{l:consistencyEndGBS}
\end{proofL}

\begin{lemma}\label{l:writeTermination}
Let us consider the algorithm in Fig. \ref{fig:server}-\ref{fig:client}. If a correct client invokes a {\sf write}$()$ operation, it eventually returns from that operation. 
\end{lemma}

\begin{proofL}
	The proof simply follows by considering that, for a ${\sf write}()$ operation invoked at some round $r$,  the ${\sf write\_confirmation}$ is generated by the client at the end of the same round just checking the value of the variables initialized at the beginning of $r$.
	\renewcommand{\toto}{l:writeTermination}
\end{proofL}
	
\begin{lemma}\label{l:readTermination}

Let $\alpha_{Mi}$ and $\beta_{Mi}$ be the parameters for each of the $4$ failure models Mi as reported in Table \ref{t:parameters} and used by the algorithm in Fig. \ref{fig:server}-\ref{fig:client}.
Let $n > \alpha_{Mi} f$ for each failure model Mi considered. 
If a correct client invokes a ${\sf read}()$ operation, it eventually returns from that operation. 
\end{lemma}

\begin{proofL}
Let $c_j$ be a client invoking a ${\sf read}()$ operation at some time $t$. 
When this happens, $c_j$ flags that a ${\sf read}()$ operation is starting and prepares a {\sc read}$()$ message to send at the beginning of the next {\em send} phase at round $r$. 
When $c_j$ sends such {\sc read}$()$ message, it updates its $op\_start_j$ variable to $r$ and it returns from the ${\sf read}()$ operation at round $r+1$ if and only if it has at least $n-\beta f$ occurrences of the same value in the $replies_j$ set.
Such $replies_j$ is initially empty (it has been emptied at the end of the previous ${\sf read}()$ operation) and it is filled in when $c_j$ receives a {\sc reply}$()$ message (line \ref{c11}) that is sent at least by non-faulty servers when they receive a {\sc read}$()$ message.

In particular, the {\sc read}$()$ message sent by $c_j$ will be delivered by servers during the receiving phase of round $r$. When this happens, any non-faulty server will execute line \ref{s18} in Figure \ref{fig:server} and will store the identifier of $c_j$ in order to send a reply at the beginning of the next round $r+1$. 
Due to Lemma \ref{l:consistencyEndGBS}, at the end of round $r$, at least $n-f$ non-faulty servers will store the same value $v$.
Let us note that, during the send phase of round $r+1$, $x$ of such servers may become faulty.
Thus, $c_j$ will find a value satisfying the condition in line \ref{c21} if and only if $n-f-x \ge n-\beta f$.
Considering that $x\le f$ for all models but Burhman's one where $x=0$, we have that the condition is always true and the claim follows.

	\renewcommand{\toto}{l:readTermination}
\end{proofL}

\begin{theorem}[Termination]\label{t:termination}
	If a correct client invokes an operation, it eventually returns from that operation. 
\end{theorem}

\begin{proofT}
	It follows direclty from Lemma \ref{l:writeTermination} and Lemma \ref{l:readTermination}.
	\renewcommand{\toto}{t:termination}
\end{proofT}

\begin{theorem}[Validity]\label{t:validity}
Let $\alpha_{Mi}$ and $\beta_{Mi}$ be the parameters for each of the $4$ failure models Mi as reported in Table \ref{t:parameters} and used by the algorithm in Fig. \ref{fig:server}-\ref{fig:client}.
Let $n > \alpha_{Mi} f$ for each failure model, Mi, considered. 
Any ${\sf read}()$ operation returns the last value written before its invocation, or a value written by a concurrent ${\sf write}()$ operation. 
\end{theorem}

\begin{proofT}
Without loss of generality, let us consider the first ${\sf write}(v)$ operation $op_W$ and the first ${\sf read}()$ operation $op_R$. 
Three cases may happen: (i) $op_R \prec op_W$, (ii) $op_W \prec op_R$ and (iii) $op_W ~|| ~op_R$.
Let us note that $op_r$ spans over two rounds: in the first one it sends the {\sc read}$()$ message and in the second one it collects replies.

	\begin{itemize}
		\item {\bf Case (i):} $\mathbf{op_R \prec op_W}$. This case follows directly from Lemma \ref{l:consistencyEndGBS} considering that (i) at the end of the first round of $op_r$ (i.e., $r_1$) at least $n-f$ correct processes have the same initial value $v=\bot$, (ii) while moving to the second round of $op_R$, at most $x$ processes can get faulty (with $x\le f$ for models M1-M3 and $x=0$ for M4), (iii) $n-f-x \ge n- \beta_{Mi}f$ (i.e. $\beta_{Mi}f \ge f+x$) for each model (i.e. there will always be enough replies from correct servers to select a value) and (iv) $n- \beta_{Mi}f > f$ (i.e. $(\alpha_{Mi}-\beta_{Mi})f +1 > f$) for each model.
It follows that faulty processes cannot force the client to select a wrong value.\\
		
		\item  {\bf Case (ii):} $\mathbf{op_W \prec op_R}$. Let $r$ be the round at which $op_W$ terminates and let $r+1$ be the round at which $op_R$ is invoked.
		
		Due to Lemma \ref{l:consistencyEndGBS}, at round $r+2$ there are enough occurrences (at least $n-\beta f$) of the last written value $v$. So, applying the same reasoning of case (i) the claim follows.\\	
		
		\item {\bf Case (iii):} $\mathbf{op_W ~|| ~op_R}$. Let us note that a ${\sf read}()$ operation spans two rounds, i.e., the round of the request $r_{req}$ and the round of the reply $r_{reply}$. So, let us consider them separately.
		
		\begin{itemize}
			\item {\bf Case (iii-a):} $op_W$ is concurrent with $op_R$ during $r_{req}$. In that case the value $v$ is delivered to correct server at the end of $r_{req}$. Due to Lemma \ref{l:consistencyEndGBS}, at the end of $r_{req}$ at least $n-f$ correct servers store the new written value $v$, we fall down into case (ii) and the claim follows.\\
			\item {\bf Case (iii-b):} $op_W$ is concurrent with $op_R$ during $r_{replay}$. Since, in every round, the send phase is executed before the receive phase, it follows that at least all the correct servers will reply with the value written before the invocation of the ${\sf write}()$ operation, we fall down into case (i) and the claim follows.
		\end{itemize}	
	\end{itemize}
	
\renewcommand{\toto}{t:validity}
\end{proofT}

\begin{theorem}[Ordering]\label{t:ordering} There exists a total order $S$ of ${\sf read}()$ and ${\sf write}()$ operations such (i) if $op \prec op'$ then $op$ appears before $op'$ in $S$ and (ii) any ${\sf read}()$ operation returns the value $v$ written by the last ${\sf write}()$ preceding it in $S$.
\end{theorem}

\begin{figure}
	\begin{tikzpicture}
	
	\draw[->] (-.3, 0) -- (-.2,0) (-.1,0) -- (7,0);\node[] at (6.9,.2) {$r$};
	
	\filldraw[fill=blue!20!white, draw=black] (0,0) rectangle (3,.4);
	\filldraw[fill=red!20!white, draw=black] (0,0) rectangle (1.5,-.4);
	\draw (0,-.5) -- (0,.5) (1.5,-.5) -- (1.5,.5);
	\draw[thick] (3,-.5) -- (3,.5);
	\filldraw[fill=green!20!white, draw=black] (3,0) rectangle (6,.4);
	\filldraw[fill=red!20!white, draw=black] (3,0) rectangle (4.5,-.4);
	\draw  (4.5,-.5) -- (4.5,.5);
	\draw  (6,-.5) -- (6,.5);
	
	\node[] at (.75,-.25) {$r_{v_1}$}; \node[] at (.75,.2) {$r_{R1_{req}}$}; \node[] at (2.3,.2) {$r_{R1_{reply}}$};
	\node[] at (1.65,.7) {$op_{R1}$}; \node[] at (.8,-.7) {$op_{W(v_1)}$};
	\node[] at (3.75,-.25) {$r_{v_2}$}; \node[] at (3.75,.2) {$r_{R2_{req}}$}; \node[] at (5.3,.2) {$r_{R2_{reply}}$};
	\node[] at (4.65,.7) {$op_{R2}$}; \node[] at (3.75,-.7) {$op_{W(v_2)}$};
	
	\end{tikzpicture}
	\caption{A general scenario which show how two subsequent ${\sc read}()$ operations $op_{R1}$ and $op_{R2}$ can not return respectively $v_1$ and $v_2$ if $v_2$ has been written before $v_1$.}
	\label{fig:scenarioOrdering}
\end{figure}
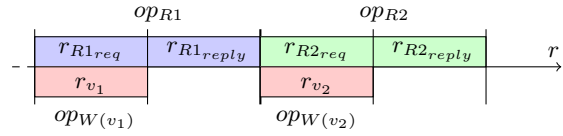

\begin{proofL}
	Consider two ${\sf read}()$ operations, $op_{R1}$ and $op_{R2}$ returning respectively $v_1$ and $v_2$ (with $v_1 \neq v_2$) such that $op_{R1} \prec op_{R2}$.
	Note that if $op_{R1}$ returns $v_1$, it follows that there exists a ${\sf write}(v_1)$ operation,  $op_{W(v_1)}$ concurrent or preceding it in $S$. 
	Suppose by contradiction that $op_{W(v_2)} \prec op_{W(v_1)}$. 
	Recall that each ${\sf read}()$ operation spans over two rounds and call the first $r_{req}$ and the second $r_{reply}$.
	Since $op_{R1}$ returns $v_1$ this means that $v_1$ has been stored by servers at latest during $r_{req}$ of $op_{R1}$; let us call it $r_{R1req}$. The same holds for $op_{R2}$: $v_2$ has been written at most during $r_{R2req}$ of $op_{R2}$. 
	Since $op_{R2}$ follows $op_{R1}$ then $r_{R1req}<r_{R2req}$. However, which is a contradiction to respect the assumption of $r_{v1}>r_{v2}$ (a general scenario is depicted in Fig.\ref{fig:scenarioOrdering}).
	\renewcommand{\toto}{t:ordering}
\end{proofL}

\begin{theorem}\label{th:Garay}
	Let $\mathcal{A}_{Areg}$ be the algorithm in Fig. \ref{fig:server}-\ref{fig:client} and let $n > \alpha f$.
	If $\alpha = 3$ and $\beta = 2$ then $\mathcal{A}_{Areg}$ implements a MWMR Atomic register in the Garay's model.
	
\end{theorem}

\begin{proofT}
	It follows directly from Theorem \ref{t:termination}, \ref{t:validity} and \ref{t:ordering}.
	\renewcommand{\toto}{th:Garay}	
\end{proofT}

\begin{theorem}\label{th:Bonnet}
Let $\mathcal{A}_{Areg}$ be the algorithm in Fig. \ref{fig:server}-\ref{fig:client} and let $n > \alpha f$.
	If $\alpha = 4$ and $\beta = 2$ then $\mathcal{A}_{Areg}$ implements a MWMR Atomic register in the Bonnet's model.
\end{theorem}

\begin{proofT}
	It follows directly from Theorem \ref{t:termination}, \ref{t:validity} and \ref{t:ordering}.
	\renewcommand{\toto}{th:Bonnet}	
\end{proofT}

\begin{theorem}\label{th:Sasaki}
Let $\mathcal{A}_{Areg}$ be the algorithm in Fig. \ref{fig:server}-\ref{fig:client} and let $n > \alpha f$.
	If $\alpha = 4$ and $\beta = 2$ then $\mathcal{A}_{Areg}$ implements a MWMR Atomic register in the Sasaki's model.
\end{theorem}

\begin{proofT}
	It follows directly from Theorem \ref{t:termination}, \ref{t:validity} and \ref{t:ordering}.
	\renewcommand{\toto}{th:Sasaki}	
\end{proofT}

\begin{theorem}\label{th:Burhman}
Let $\mathcal{A}_{Areg}$ be the algorithm in Fig. \ref{fig:server}-\ref{fig:client} and let $n > \alpha f$.
	If $\alpha = 2$ and $\beta = 1$ then $\mathcal{A}_{Areg}$ implements a MWMR Atomic register in the Burhman's model.\end{theorem}

\begin{proofT}
	It follows directly from Theorem \ref{t:termination}, \ref{t:validity} and \ref{t:ordering}.
	\renewcommand{\toto}{th:Burhman}	
\end{proofT}

\section{Conclusion}\label{sec:conclusions}
This paper addressed the first implementation of a  {multi-writer multi-reader} atomic register tolerant to mobile Byzantine agents altogether with upper bounds on the number of faulty processes. We investigate four models of mobile Byzantines in round-based synchronous systems: the model of Garay {\it et al.} \cite{Garay+1994}, where nodes have the capability to detect an infection and clean their state after the Byzantine agent leaves the node; the models of Sasaki  \emph{et al.}  \cite{Sasaki+2013} and Bonnet  {\it et al.} \cite{BDNP14}, where infected nodes may execute their code with a corrupted state even though the mobile agent is not anymore located at the node and finally, the model of Buhrman \emph{et al.} \cite{Garay+95+ORA} where Byzantines move are tight to messages and move during the send phase. As for the case of the agreement problem (benchmark already investigated in all these models) our study shows that the atomic registers cannot be implemented using the static bounds on the number of faulty processes. That is, we prove that in the Garay's model atomic registers can be implemented provided that in each round the number of Byzantine nodes (nodes occupied by a Byzantine agent), $f$, is less than $n/3$ where $n$ is the number of correct nodes in that round while in the Bonnet's and Sasaki's models the number of Byzantine nodes $f$ is less than $n/4$.  Finally, for the case of Buhrman's model we show that $f$ should be less than $n/2$. 
Our study can be extended in several directions (here after we mention only two of them). First, an interesting issue is to investigate the storage problem in the round-free synchronous and furthermore in the asynchronous settings. We conjecture that in these models the bounds on the faulty processes are different from the round-base case. Secondly,  our study advocates in favor of  revisiting other building blocks of distributed computing in these settings (e.g. quorums, k-set agreement, synchronization etc).  In all these cases we conjecture lower and upper bounds different from the static case. 

\bibliographystyle{splncs}
\bibliography{references,references1,biblio}

\end{document}